*Review article*

# A quarter century of astrophysics with Japan

Philip Yock

Department of Physics, University of Auckland, PO Box 92109, Auckland 1142

On 23 February 1987 a supernova (exploding star) was observed in the Large Cloud of Magellan, the brightest supernova in 400 years. It spurred the commencement of collaborative research in astrophysics between Japan and New Zealand that is still ongoing after 25 years. The initial aim of the two countries was to search for evidence of cosmic rays being emitted by the supernova in a project named JANZOS. A large cosmic ray detector was installed near the summit of the Black Birch range in Marlborough to monitor the supernova but, after seven years of operations, the results proved to be negative. In 1994 a second phase of research was commenced, this time at the Mt John University Observatory in Canterbury under the emblem MOA. The aim of the MOA project is to study dark matter and extrasolar planets using a novel 'gravitational microlensing' technique. A 1.8 m telescope was built at the observatory, and several planets were subsequently found. Some have exotic properties, the most exotic being a new class of 'free-floating' planets that appear not to be orbiting stars. They are estimated to outnumber stars by about two to one. In this article the work carried out by the JANZOS and MOA collaborations is reviewed, and plans for the future are outlined.

## Introduction

In February 2012, Japan and New Zealand passed a quarter century milestone for jointly conducting research in astrophysics. The research began on 23 February 1987 with the observation of a supernova (exploding star) in the Large Cloud of Magellan, the brightest in 400 years. The supernova, known as SN1987A, offered unique opportunities for research, and it triggered the commencement of collaborative research in astrophysics between Japan and New Zealand. The supernova was monitored from a high-altitude site in the Black Birch range in Marlborough from 1987 to 1994. In the process, a productive working relationship between the partner countries developed.

In 1994 a second joint venture named MOA (Microlensing Observations in Astrophysics) was launched. This project, which is based at the Mt John University Observatory in Canterbury, uses a 'gravitational microlensing' technique that depends on

Correspondence: p.yock@auckland.ac.nz

Einstein's theory of gravity to hunt for dark matter and extrasolar planets. The MOA project is still continuing. Indeed, after 18 years, it is still in a phase of expansion. Recent publications include a growing list of participating astronomers, presently totalling about 150 from some 20 countries and 50 institutions.

A few members of the original partnership from both Japan and New Zealand are still actively engaged in the research. To mark the quarter century milestone it seemed opportune to chronicle the work done together in a non-specialist article.

## The JANZOS project

Supernova SN1987A was discovered independently by Ian Shelton from Chile and Albert Jones from New Zealand. Its proximity in the Large Cloud of Magellan offered unique opportunities for research. Within days of the discovery, I received a cable from a Japanese colleague, Yasushi Muraki, then at the University of Tokyo, requesting assistance to install a cosmic ray detector at high altitude in New Zealand to see if the supernova was emitting cosmic rays. These are high-energy particles, mostly protons, that are trapped within the galaxy by its magnetic field. Their origin is unknown, although the remnants of supernovae have long been thought to be a contributing source. Their energies extend to more than $10^{20}$ eV, some eight orders of magnitude greater than the energies of protons accelerated by the Large Hadron Collider.

The unique opportunity offered by the request from Japan could hardly be turned down, and the setup shown in Figure 1 was installed as rapidly as possible at an altitude of 1600 m in the Black Birch range in Marlborough during the winter of 1987. It covered five hectares and was sensitive to cosmic rays and gamma rays with energies in the $10^{12}$ to $10^{15}$ eV range. The equipment was installed above a field station that was operated by Carter Observatory at Black Birch at the time, and also above an astrometric observatory that was operated by the US Navy then. Both these institutions, together with the University of Tasmania and the Ministry of Works, provided much appreciated assistance, and the work proceeded under the banner

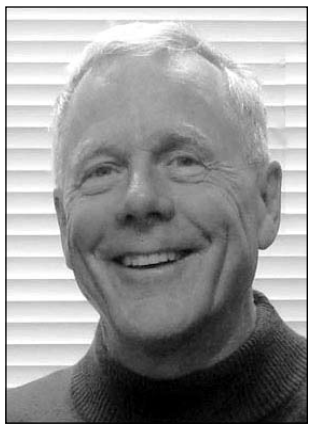

**Philip Yock** is Associate Professor of Physics at the University of Auckland. He is a Fellow of the Royal Astronomical Society of New Zealand and a Member of the New Zealand Order of Merit. Together with Professor Yasushi Muraki of Konan University he co-founded the JANZOS and MOA projects with Japan in 1987 and 1994 respectively. He currently works on the MOA project and on particle physics.

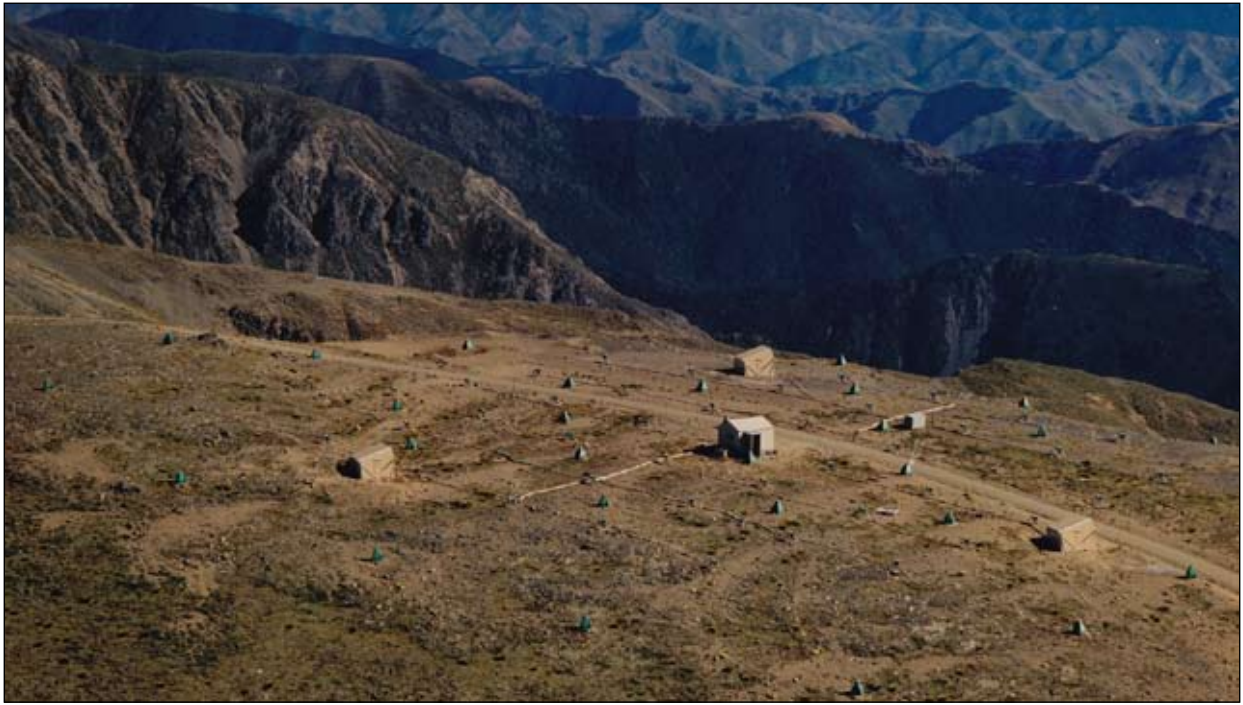

*Figure 1. JANZOS cosmic ray detector in the Black Birch range used to hunt for gamma rays from SN1987A in the Large Magellanic Cloud from 1987 to 1994 by Japan, Australia, and New Zealand.*

Japan Australia New Zealand Observation of Supernova, or JANZOS for short.

The specific aim of the JANZOS project was to search for gamma rays coming from the direction of SN1987A. Since gamma rays are undeflected by the galactic magnetic field, a positive detection would have indicated that cosmic rays were being produced by the supernova, and undergoing interactions in its ejecta in which gamma rays were being produced.

The 76 green pyramidal enclosures shown in Figure 1 housed scintillation detectors whose signals were routed to the central electronics hut. The scintillators acted as an all-sky 24-hour camera with a state-of-the-art directional sensitivity of 1° at $10^{14}$ to $10^{15}$ eV energies. This was confirmed by verifying that shadows cast by the Moon and the Sun were present in the data. The three enclosures with sloping, retractable roofs enclosed Cerenkov telescopes that were sensitive in the $10^{12}$ to $10^{13}$ eV range. They operated in hours of darkness only.

SN1987A and other potential sources of high-energy radiation in the southern sky were monitored from 1987 to 1994 with the above setup but, alas, the results proved to be negative. The mystery of the origin of the cosmic radiation remained unsolved, as it is today (Raymond 2009; Clery 2012). Our measurements were reported in a series of publications during the 1980s and 1990s (Bond et al. 1988a,b, 1989; Allen et al. 1993a,b,c,d, 1995; Abe et al. 1999a).

## The birth of the MOA project

Despite the physically challenging qualities of the site at Black Birch, which often included snow or sleet and winds too strong to stand in, many excellent theses were written by students from both Japan and New Zealand on the JANZOS project, and a positive working relationship was formed between the countries. The inclement weather presented a challenge that neither country was willing to be first to give in to.

However, in 1994 a joint decision was made to re-direct our skills in a new direction, notably the hunt for dark matter and extra-solar planets using the then-new technique of 'gravitational microlensing'. Like the JANZOS project, this required a southern location. We rebranded ourselves 'Microlensing Observations in Astrophysics' or MOA, and moved from wind-swept Black Birch to sunny Mt John in Canterbury. Thus was born the MOA project.

It is noteworthy that the study of extra-solar planets was far from fashionable in 1994. Only two extra-solar planets were known at the time, and these orbited a pulsar. They shared little in common with the extra-solar planets orbiting normal stars that we sought. Our quarry was closer to that envisaged by the early astronomers such as Newton who speculated in Principia that 'if the fixed stars are the centres of similar (planetary) systems, they will all be constructed according to a similar design' (Newton, 1713).

The first planet to be found in orbit around a normal star was not reported until 1995 (Mayor & Queloz 1995), just one year after the commencement of MOA. In contrast, the hunt for dark matter was a fashionable, but unsolved, line of research in 1994, as it is today.

## Gravitational microlensing

The phenomenon of gravitational microlensing is caused by the bending of light in the gravitational field of a massive object as predicted by Einstein (1936). It is depicted in Figure 2. This shows an 'Einstein ring' being formed when two distant stars are collinear as seen from Earth.

In 1936 Einstein had predicted that the phenomenon of microlensing would never be observed. His opinion was based on the observational powers of the lone astronomer of his day observing stars one at a time. Today's telescopes have wide fields-of-view, and large electronic cameras coupled to equally large computers. They record images of millions of stars simultaneously. This permits rare alignments of stars to be found and monitored whilst they persist.

For typical stars at galactic distances, the size of the Einstein ring is comparable to that of the asteroid belt in our solar system. If planets orbit the nearer star, their gravitational fields can distort the Einstein ring appreciably, and hence betray their presence (Liebes 1964, Mao & Paczynski 1991). The phenomenon is best observed from the southern hemisphere because the densest stellar fields are in the centre of the Galaxy, and this lies in the southern sky (in Sagittarius).

The process of planet detection by microlensing may be compared to Rutherford's experiment on the scattering of α-particles by gold atoms. In Rutherford's experiment the α-particles were deflected by the gold nucleus, but orbiting electrons played a secondary role by decelerating the α-particles slightly. In microlensing, photons are deflected primarily by the star at the centre of the lens system, but orbiting planets play a secondary role by deflecting the photons slightly.

In contrast to other techniques used to hunt extra-solar planets, the microlensing method does not rely on the emission of light by stars hosting planets. It is possible to detect planets orbiting very faint stars, such as red dwarfs or even brown dwarfs. It is also possible to detect darker objects such as lone planets (not orbiting stars) or black holes – hence the motivation for searching for dark matter.

MOA uses the two telescopes shown in Figure 3. For the first ten years only the smaller telescope (0.6 m aperture) on the right of the photograph was available. This was generously supplied by Canterbury University, thanks largely to the efforts of John Hearnshaw. It was modified at the University of Auckland for microlensing with the installation of wide-angle optics and a computerised drive system, and equipped with CCD (charged couple device) cameras supplied by Japan.

The larger (1.8 m) telescope was installed in 2004. It was supplied by Japan thanks largely to the efforts of Yasushi Muraki. The optical design was supplied by Andrew Rakich, a graduate of Canterbury University, and construction was carried out by a smallish family business in Kyoto headed by Yuji Nishimura. The CCD camera for this telescope was built at Nagoya University under the direction of Takashi Sako. Further information on both telescopes is given in Yock (2006).

## The gravitational microlensing community

The MOA group is not alone in hunting planets by microlensing. Indeed, the field has flourished, thanks to the formation of a sizable community utilising the technique.

Before the MOA group got under way in New Zealand, a Polish group named OGLE[1] commenced observations from Chile. Over the years the OGLE group has steadily upgraded their setup, as has MOA. The MOA and OGLE telescopes complement one another, thanks to their longitudinal separation. Between them, they monitor about 100 million stars in the centre of the Milky Way several times per (clear) night during southern winters. About 1000 examples of microlensing are found annually.

The brightest of these are selected for 'follow-up' observations by networks[2] of telescopes named PLANET, MicroFUN, RoboNET and MiNDSTEp. They operate telescopes in New Zealand, Australia, South Africa, Chile, Hawaii, continental USA, Israel, Antarctica, Tahiti, and the Canary Islands.

In addition to the above networks, observations of gravitational microlensing events are made frequently by the Very Large Telescopes at the European Southern Observatory in Chile and by the Hubble Space Telescope. Another space telescope known as 'Deep Impact' has been used recently (Muraki et al. 2011). Deep Impact is now 1.6 AU from Earth (1 AU = the Sun–Earth distance) and its images provide stereoscopic views of microlensing events when combined with Earth-based images.

With this armada of telescopes, totalling more than twenty, there is sufficient redundancy to record the peaks of most of the brightest microlensing events continuously, even allowing for cloudy weather at some sites.

## Main results obtained by the microlensing community

### Cool planets

Ground-breaking results on planets orbiting stars at 1–5 AU from their host stars have been obtained by the microlensing community. The 1–5 AU range includes the 'snow-lines' around stars where solid ices form in proto-planetary disks and enhanced formation of ice-giant planets similar to Uranus and Neptune occurs (Ida & Lin 2004; Papaloizou & Terquem 2006). It also marks the distance where gas giants like Jupiter and Saturn form, by accreting hydrogen from the proto-planetary nebula onto ice-giant cores, but other formation mechanisms for these planets are also possible (Boss 1997).

Planets beyond the snowline are not easily detected by other techniques, and the microlensing community has cornered the market on them for smaller-sized planets, as shown in Figure 4.

To convert plots such as Figure 4 to planetary abundances one needs to fold in the detection efficiencies of the various techniques. When this is done, one discovers that the planets being found by microlensing are abundant.

Using data acquired by the microlensing community as a whole, the PLANET group (Cassan et al., 2012) recently estimated that about 7% of stars host Jupiter-like planets, approximately 50% host Neptune-like planets, and very roughly 60% host smaller planets with masses some five to ten times that of Earth. Although the PLANET group was unable to determine the abundance of Earth-like planets, they were able to conclude that stars are orbited by planets as a rule, as anticipated by Newton (1713).

### Free-floating planets

However, not all the findings have followed expectation. The MOA group, with assistance from the OGLE group, recently found evidence for a large population of planet-like objects that was unexpected. These are Jupiter-mass objects but they appear not to be orbiting stars. Instead, they roam interstellar space on their own. They are sometimes referred to as 'free-floating' or 'rogue' planets (Sumi et al. 2011).

---

[1] OGLE Collaboration: http://ogle.astrouw.edu.pl/

[2] PLANET Collaboration: planet.iap.fr
MicroFUN Collaboration: www.astronomy.ohio-state.edu/~microfun
RoboNET Collaboration: robonet.lcogt.net
MiNDSTEp Collaboration: www.mindstep.science.org

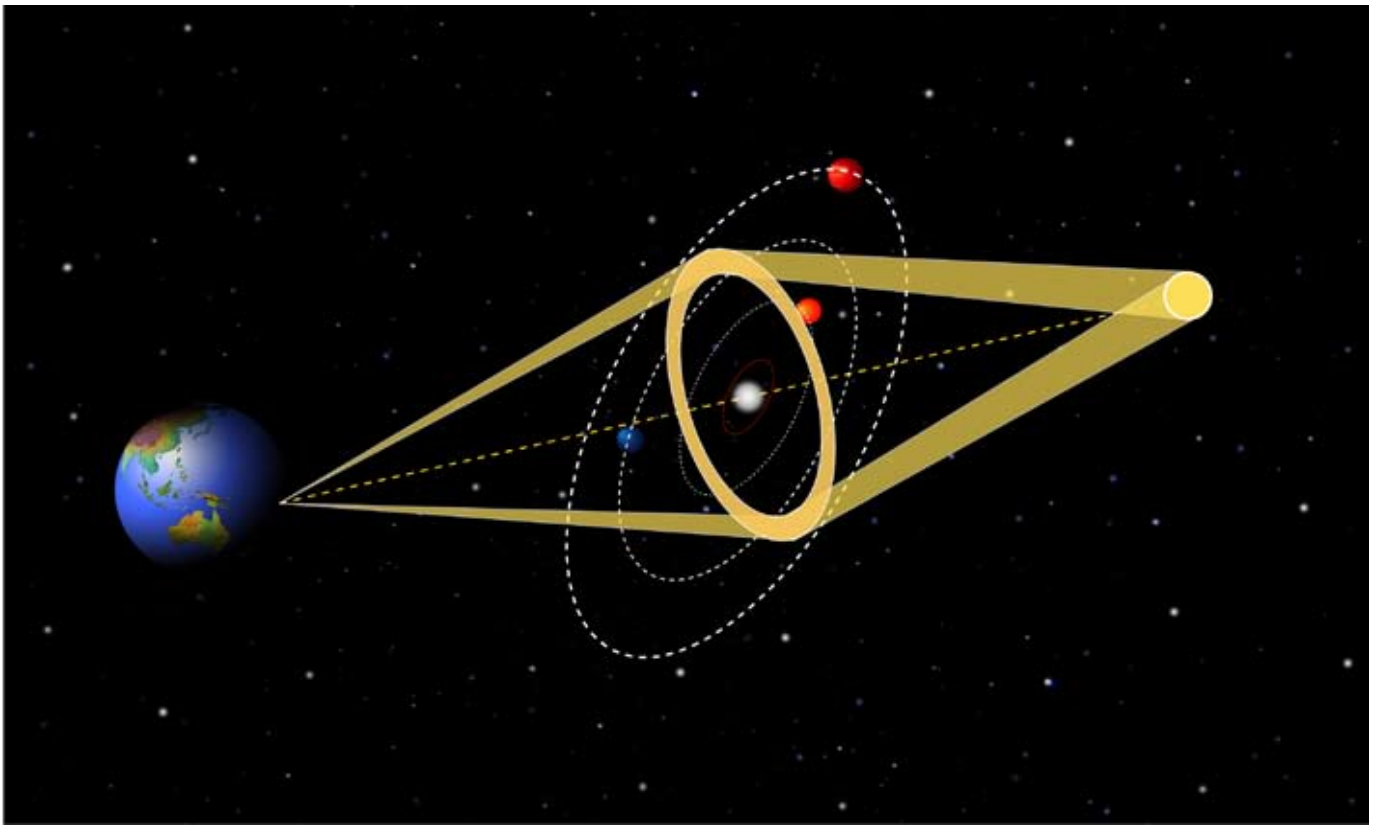

*Figure 2. Formation of an Einstein ring when light from a distant 'source' star is bent by the gravitational field of a closer but collinear 'lens' star. Similar lensing is caused by lone planets in interstellar space, but they produce smaller Einstein rings because they are less massive.*

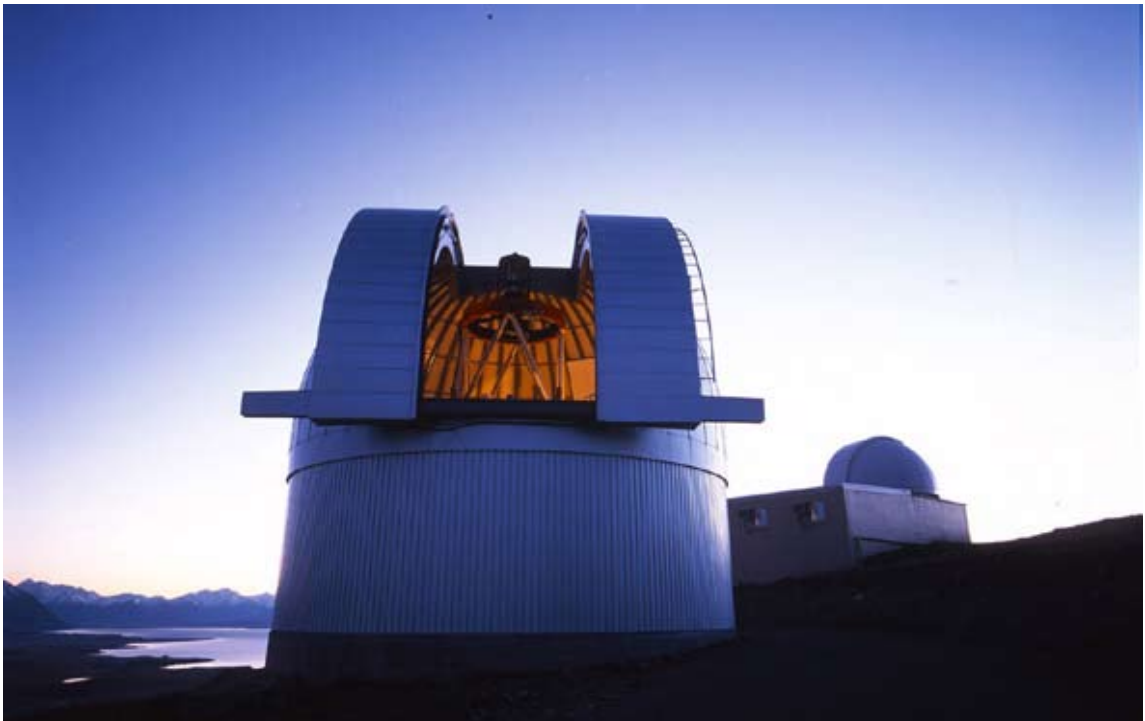

*Figure 3. Telescopes (1.8 m and 0.6 m) used by the MOA group at the Mt John University Observatory. Lake Alexandrina and the Southern Alps are in the background.*

Free-floating planets were found in microlensing events with Einstein rings approximately twenty times smaller than those caused by stars. Such rings are formed by objects of similar mass to Jupiter. The initial discovery of the anomalous events was made at Nagoya University, by then graduate student Koki Kamiya working under the direction of Taka Sumi.

The abundance of free-floating planets was estimated from the frequency of events with anomalously small Einstein rings to be a staggering two times higher than the abundance of normal stars in the Galaxy. Although we cannot see them with the unaided eye, the data indicate twice as many free-floating planets between the stars as the actual stars we see! The nearest ones may be only a few light years away.

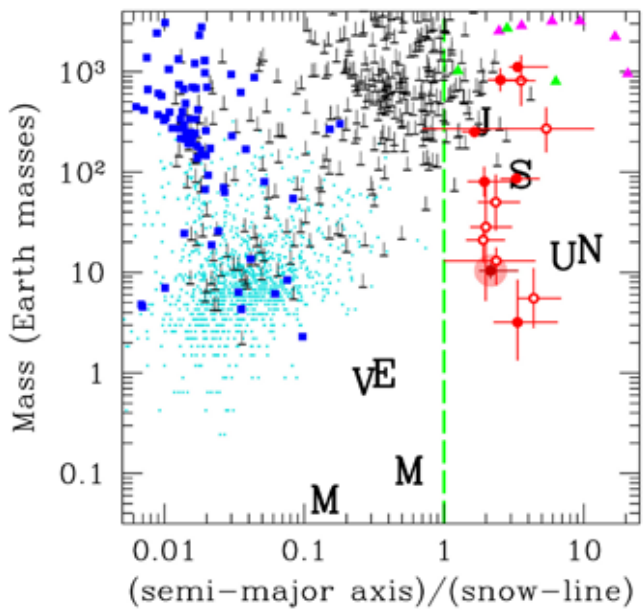

*Figure 4. Masses and semi-major axes scaled to the snowline of known planets as at December 2011. Discoveries by other techniques are shown in other colours. The Kepler space mission by NASA revealed the cluster of planets shown in cyan. The planets of our solar system are denoted by letters, with Mercury the lightest and closest to the Sun (Muraki et al. 2011).*

The origin of these unexpected objects is presently unknown. They may have been born as members of planetary systems and subsequently ejected in gravitational planet–planet or planet–star interactions (Sumi et al. 2011), or they may have been born as lone Jupiter-sized objects in which case they might be better characterised as 'misfit' or 'failed' stars with too little mass to support nuclear burning (Grifantini 2012). Either way, further observations are definitely needed. An artist's depiction of one of them appears in Figure 5.

Free-floating planets may be able to be detected using other techniques. A search was begun recently by graduate student John Bray at the University of Auckland using data from the WISE space mission. This mission scanned the entire sky at infrared wavelengths and it could reveal free-floating planets if some are located within a few light years of us. If nearby ones can be found this way, they will be able to be examined in fine detail with NASA's upcoming James Webb Space Telescope, the successor to the Hubble Space Telescope.

*Figure 5. Artist's depiction (not to scale) of free-floating planet MOA-ip-10 at the infrared wavelengths at which it might be radiating. The two arcs are the magnified images of a background star caused by the planet when it was in near-perfect alignment with a background star. Background stars of the Milky Way are included at infrared wavelengths. [Credit: Jon Lomberg]*

Unfortunately, free-floating planets found by microlensing are too distant to be observed by the James Webb Space Telescope. Also, the analyses of most of the observations made to date are statistical in part. The first example that was free of statistical arguments was reported earlier this year by MSc student Matthew Freeman of the University of Auckland (Freeman 2012).

## Future prospects

Future prospects for microlensing are excellent. The MOA and OGLE telescopes will be joined in the next two years by three comparable Korean telescopes in Australia, South Africa and Chile (Kim et al. 2011). Also, MOA has plans (yet to be funded) for another telescope to be built in Namibia. With these additional telescopes, the repetition rate for monitoring stars in the centre of the Galaxy will be reduced to about 10 minutes.

In addition, an educational organisation known as Las Cumbres Observatory Global Telescope (LCOGT) will shortly expand their suite of follow-up telescopes with six additional 1 m telescopes in South Africa and Chile (Brown et al. 2012).

With these new telescopes, one can anticipate a significantly increased detection rate of planets, increased understanding of planetary formation processes, more information on free-floating planets, and hopefully some information on the abundance of planets that could host life.

On the question of habitability, one of course needs to allow for the possible existence of life forms quite distinct from those on Earth. It has been suggested, for example, that free-floating planets could be warmed to habitable temperatures from the combination of heating by internal radioactivity and blanketing by a thick atmosphere of hydrogen (Stevenson 1999).

At least three dedicated space missions are under consideration for future observations. In 2010 the US National Academy of Sciences recommended top priority for a US$1.6 billion space mission for the next decade named WFIRST to study dark energy and extrasolar planets by the microlensing technique (Blandford et al. 2010) but cost overruns of the James Webb Space Telescope have placed this mission in jeopardy. Interestingly, however, two spare space telescope mirrors miraculously turned up recently (Overbye 2012). They were produced for the US military but never used. One of these could be used to resurrect the WFIRST mission. Also, the European Space Agency has plans for a mission named Euclid that could include telescope time for microlensing observations similar to those planned for WFIRST (Penny et al. 2012).

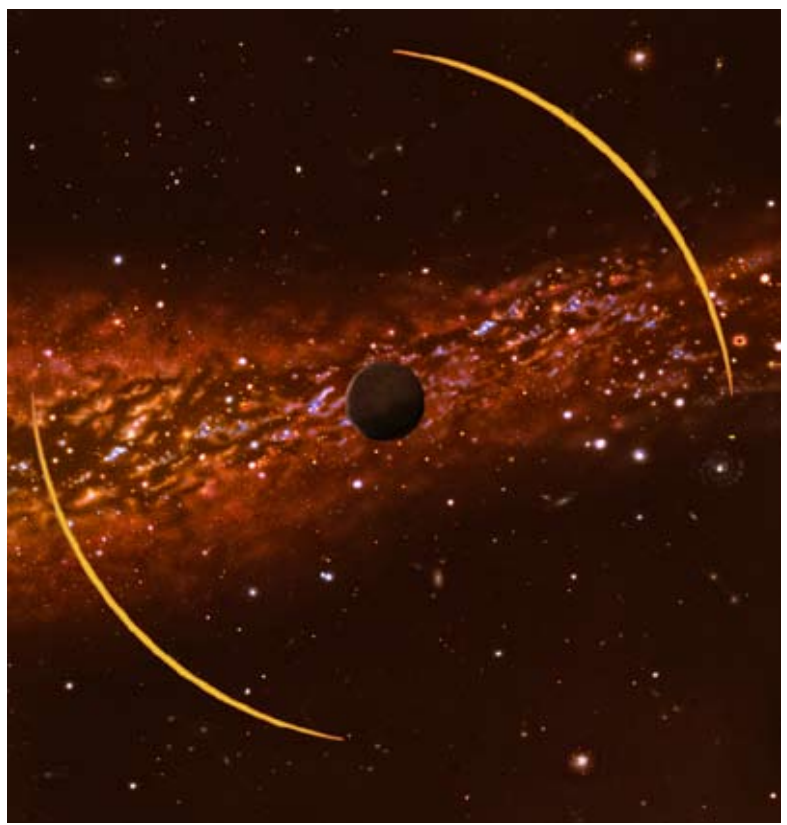

## New Zealand's contributions

One may ask how all the above came about. Did New Zealand merely host a couple of interesting international enterprises and stand by on the sidelines watching developments, or did we contribute? In answer it may be of interest to record some of the contributions made by New Zealanders, on both the JANZOS and MOA projects, especially as some colleagues expressed concern.

### *From cosmic rays to dark matter*

Amongst the publications that resulted from the JANZOS project it is noteworthy that these originated fairly equally from Japan (Bond et al. 1988a,b, 1989; Allen et al. 1993c,d) and New Zealand (Allen et al. 1993a,b, 1995; Abe et al. 1999a). The latter papers resulted from very ably written theses by students of the University of Auckland, notably Mark Conway, Peter Norris, Michael Woodhams, Ian Bond, Matthew Spencer, Grant Lythe and Anthony Daniel. The 1995 and 1999 papers reported studies of the nearest supernova remnants in the southern sky. They complemented the original papers by JANZOS on SN1987A, the brightest supernova. The result from Abe et al. (1999a) is pictured in Figure 6 (left panel).

Figure 6 also shows the first (and only) result by MOA on dark matter. The right panel is a contour plot of a five-hour exposure of the edge-on galaxy IC5249 taken by MSc student Glenn Pennycook of the University of Auckland from Mt John, in which evidence for a faint halo of red or brown dwarfs surrounding the galaxy was sought.

Red dwarfs are faint stars, only slightly brighter than their nearly invisible brown dwarf cousins. During the 1990s, brown dwarfs were thought to be a viable candidate for dark matter, and it was suggested that red dwarfs could be used as a tracer because they are similar. However, the long exposure shown above yielded a negative result (Abe et al. 1999b). A Cambridge group independently reported similar results for four other edge-on galaxies using the Hubble Space Telescope (Gilmore & Unavane 1998).

The extraordinarily dark, clear sky conditions at Mt John permitted the above five-hour exposure to be taken successfully. The darkness and clarity of the night sky were recognised recently when it was announced that the Aoraki Mackenzie region had been recognised as the world's third and largest International Dark Sky Reserve.

### *First microlensing result*

A welcome transition from null results to positive ones occurred in 1995 with a microlensing event known as MACHO-95-BLG-30. This denotes the 30$^{th}$ microlensing event found by a group called MACHO in 1995 in the direction of the bulge of stars at the centre of the Milky Way.

The MACHO group was one of the pioneering microlensing groups when observations were first undertaken (Alcock et al. 1993). They observed from Mt Stromlo in Australia, and it was a tragic loss to science when their telescope, and much of the Mt Stromlo Observatory, was destroyed by a forest fire in January 2003.

Amongst their many achievements, the MACHO group was first to realise the value of operating a worldwide network of telescopes to provide round-the-clock coverage of microlensing events. This led to a remarkable coincidence, as the principle investigator of the MACHO group, Charles Alcock, was a former student of the University of Auckland. It was therefore natural for the MACHO and MOA groups to join forces. This yielded two significant publications.

*Figure 6. (Left panel) Arrival directions of ultra-high energy cosmic rays from the direction of the nearest supernova remnant in the southern sky showing no excess in the source direction (Abe et al. 1999a).*

*(Right panel) Contour plot of an edge-on galaxy showing no evidence for dark matter composed of red or brown dwarfs surrounding the edge-on galaxy IC5249 (Abe et al. 1999b).*

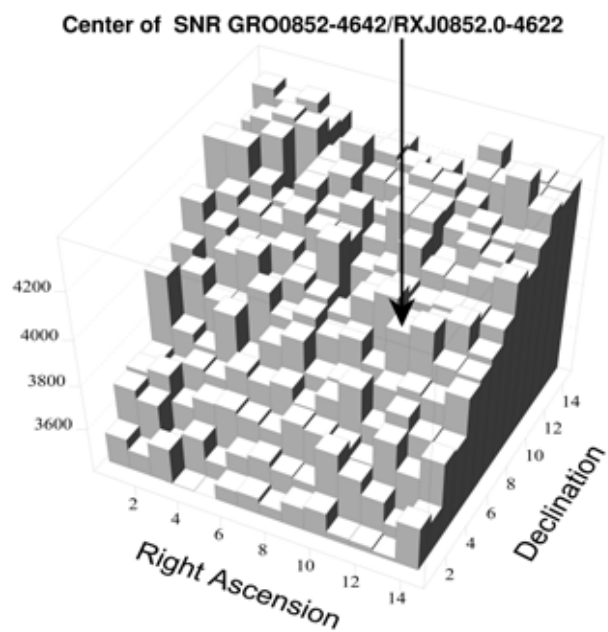
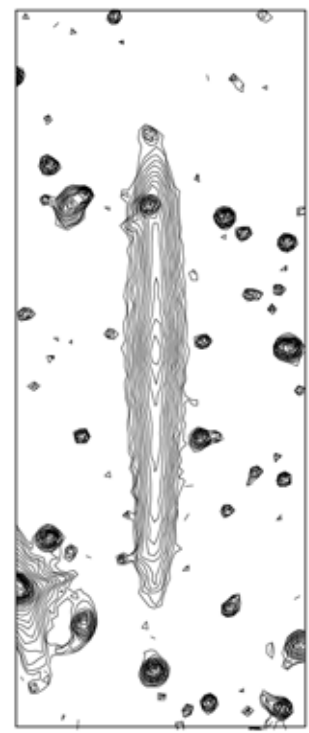

Jointly made observations of MACHO-95-BLG-30 demonstrated the potential of the microlensing technique to resolve extraordinarily small angles (Alcock et al. 1997). The angular resolution of a telescope under normal circumstances is limited by the effect of diffraction to such an extent that even the nearest stars to Earth are not resolvable with the Hubble Space Telescope. Yet the data on MACHO-95-BLG-30 showed that a star at the centre of the galaxy, some thirty thousand light years away, had been resolved (Glanz 1997). Admittedly it was a giant star, but nonetheless this represented a significant achievement. The analysis of the MOA data for this event was carried out by Denis Sullivan at Victoria University.

### *High magnification*

Three years later, joint observations were made by MACHO, MOA, and the PLANET group (mentioned previously) that demonstrated the potential of microlensing to detect planets. This occurred in event MACHO-98-BLG-35. The magnification caused by gravitational lensing in this event was the highest then recorded, approximately 80. This was fortuitous, as a US group had just pointed out that events with high magnification would be especially sensitive to the presence of planets (Griest & Safizadeh 1998).

The MOA group took on the challenge and monitored MACHO-98-BLG-35 continuously over a full night of beautifully clear weather at Mt John. The observations were made by Sachi Noda and Toshi Yanagisawa, frequent visitors to the observatory from Nagoya University where they were then PhD students. Their observations yielded possible (but not definitive) evidence for a small planet, not much larger than Earth (Rhie et al. 2000).

More importantly, MACHO-98-BLG-35 alerted members of the microlensing community to the potential of the high magnification technique. This was especially true at the University of Auckland, where a number of papers were written and conference presentations delivered on the idea (e.g. Rattenbury et al. 2002; Bond et al. 2002). These were influential. Subsequently, most of the microlensing community transferred their efforts to observing events of high magnification. This proved to be a crucial step. Most discoveries of planets by microlensing in the following decade were made in high-magnification events. Arguably, the microlensing community would not exist today had this step not been taken.

### *Crowded fields*

About the time of MACHO-98-BLG-35 another significant step forwards was taken, this time at the Institute for Astrophysics in Paris. Stellar fields at the centre of the galaxy, where nearly all observations of microlensing are made, are very crowded. The typical spacing between stars on images is only slightly larger than the typical size of a stellar image caused by the blurring effect of the Earth's atmosphere, as shown in Figure 7. Most stellar images overlap their neighbours. This presented a problem – how to extract accurate information from overlapping images?

The solution to this problem was announced by Christophe Alard at a meeting in Paris and subsequently published (Alard & Lupton 1998). With typical French flair he turned the problem into an advantage. I remember the announcement well, as I was chairman of the conference session. The potential of Alard's technique was quickly appreciated by Ian Bond, then at Auckland and Canterbury Universities, and the MOA group was first to take it on board for all measurements (Bond et al. 2001).

### *High resolution*

In 2005 the MOA group published a high-resolution image of a solar-like star some 20,000 light years away in the direction of the galactic centre. The image was obtained by microlensing (Figure 8). The angular resolution corresponds to that which would be required to read this article if it was placed on the moon, surpassing the resolution possible with the Hubble Space Telescope under normal circumstances by a factor of one million. The analysis of the data for this remarkable image was carried out by Nicholas Rattenbury, then at the University of Auckland (Rattenbury et al. 2005). The gravitational lens that provided the image was a binary star (a pair of stars orbiting their common centre-of-mass) with ideal dimensions for resolving the source star. The latter was found to be similar to the Sun.

### *Software*

As in nearly all branches of modern science, the computer has played an indelible role in the field of gravitational microlensing. Images such as that shown in Figure 7, which is only a tiny segment of a complete image, of which 400 may be taken in a single night, quickly require a powerful computing system to keep track of all the information contained on them. Ian Bond, now at Massey University, is a world leader in this sphere.

Sophisticated codes are also required to interpret the so-called 'light curves' that occur in microlensing events. These are plots of magnification versus time. Several codes have been written, overseas and in New Zealand.

Codes that are capable of handling multi-planet events with increasing sophistication were written successively by Nicholas Rattenbury, Lydia Philpott, Christine Botzler and Yvette Perrott at the University of Auckland, and by Aarno Korpela, Paul Chote and Michael Miller at Victoria University of Wellington. A typical result from these codes appears in Figure 9. In this case the data were fitted by a planetary system consisting of red dwarf at a distance of 20,000 light years being orbited by a planet of half Saturn's mass at a radius of 2.4 AU (Freeman 2010; Miyake et al. 2011).

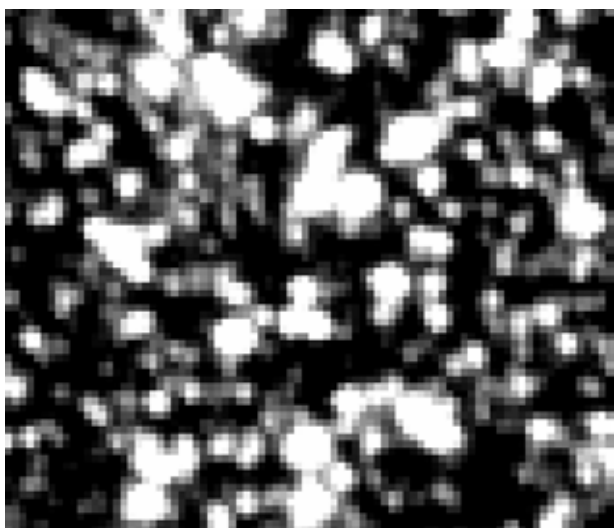

*Figure 7. A small portion measuring 80 × 80 pixels in the field-of-view of the CCD camera on the MOA telescope showing overlapping stellar images. The full field-of-view of the telescope includes 80 million pixels.*

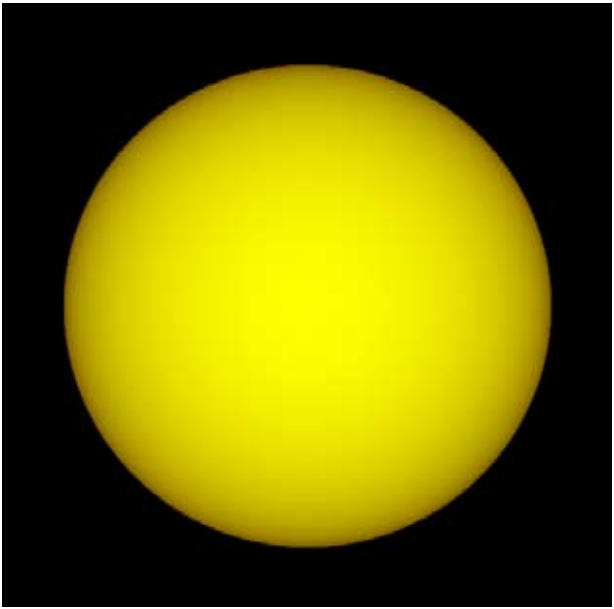

*Figure 8. High-resolution image of a distant solar-like star obtained by gravitational microlensing. The ellipticity of the star was shown to be less than 2%.*

### Earth twin

Finding an Earth twin (preferably inhabited!) is every planet hunter's dream. As Figure 4 shows, most planets found to date are hot, close-in planets orbiting their parent stars at smaller orbital radii than that of Mercury about the Sun. These are the easiest extra-solar planets to spot by most techniques, but they are probably not the most abundant. Small, warm terrestrial planets like Earth may be considerably more abundant, but they are also more difficult to detect.

A proposal for detecting small, cool terrestrial planets was made recently at the University of Auckland. This involved taking a step backwards from the high-magnification technique. When that technique was originally proposed, in 1998, it was thought that magnifications of order 100 would be the maximum that would be observable. However, in their subsequent quests for high-magnification events, the MOA and OGLE groups both found events with magnifications of about 1000 (Abe et al 2004; Dong et al. 2006). It was natural to assume that these would provide the greatest sensitivity to planets, but in fact this is not the case. Events with lower magnifications, of order 100, offer better prospects provided they are monitored with larger telescopes (Yock 2008).

It is hoped that the LCOGT group mentioned above will take up the challenge offered by the above procedure. As mentioned previously, they are currently installing six new telescopes in South Africa and Chile which, when combined with their present suite of telescopes in Hawaii, Australia and the Canary Islands, could measure the abundance of planets in the Milky Way down to Earth mass.

### *First planet in two hundred years*

New Zealanders have made other significant contributions to microlensing. Probably the most outstanding are those by so-called 'amateurs'. These are astronomers who hold down day-time jobs and observe through the nights.

The high magnifications that occur in microlensing render the relatively small telescopes used by amateurs capable of making useful observations. Grant Christie and Jennie McCormick (Figure 10), both of Auckland, led the world in this. In 2005 they observed the event OGLE-2005-BLG-71 which resulted in the first discovery of a planet by amateurs (in association with others) since William Herschel found Uranus in 1781 (Udalski et al. 2005). Since then, they and other amateurs made several such observations. New Zealand is the world leader in this regard.

*Figure 9. Magnification versus time (in days) of the peak of microlensing event MOA-2009-BLG-319. This event was monitored by 20 telescopes. The blue line is the theoretical fit with a single planet.*

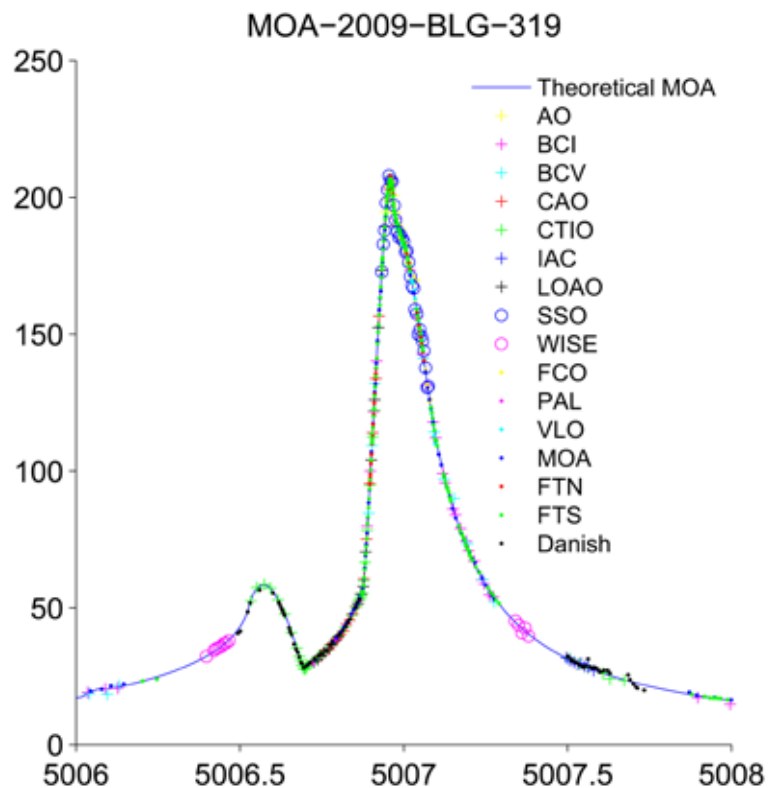

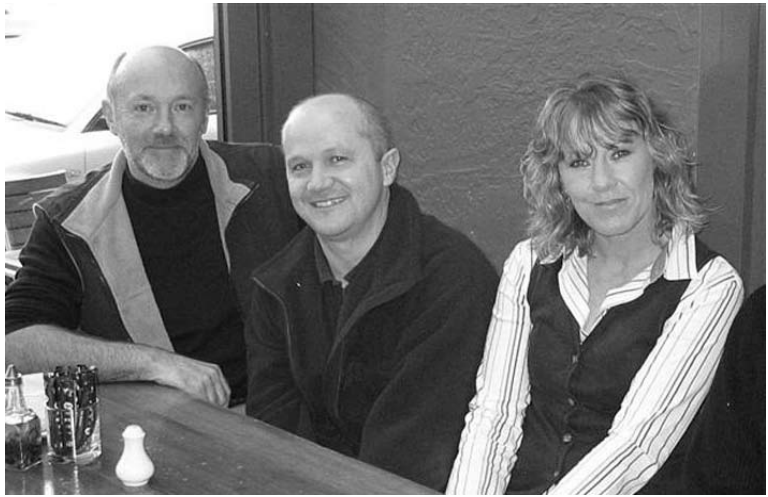

*Figure 10. Grant Christie, Ian Bond and Jennie McCormick discussing OGLE-2005-BLG-71 at Mission Bay (Auckland) after the event.*

## Conclusions

The above is a brief summary of the work done jointly by Japan and New Zealand on astrophysics during the last 25 years. Further information on the JANZOS and MOA projects is available at their websites[3]. Hopefully, readers may conclude that both countries 'discharged their duties' responsibly when the unique situation presented by the supernova of 1987 occurred. And hopefully most might agree that even Einstein would be pleased with the findings that are being made on stars and planets using his gravitational lenses.

---

[3] *JANZOS Collaboration: http://www.ritsumei.ac.jp/~morim/Janzos/index.html*
 *MOA Collaboration: http://www.phys.canterbury.ac.nz/moa/*

## Acknowledgements

A scientific partnership with Japan has been firmly and fruitfully cemented in place, and productive partnerships have also been formed with USA, Poland, France, Korea, Israel, Spain, Germany, Australia, Britain, Italy and other countries.

From a personal perspective, I have enjoyed immensely working with colleagues and students from these countries. The thought that future astronomers might be assisted by the observations is sufficient reward for the effort.

Thanks are due to the Marsden Fund of New Zealand, Earth & Sky Ltd of Tekapo, the University of Auckland and the Department of Education of Japan for financial support, and especially to Yasushi Muraki (Figure 11), for his selfless support of the New Zealand/Japan partnership for 25 years.

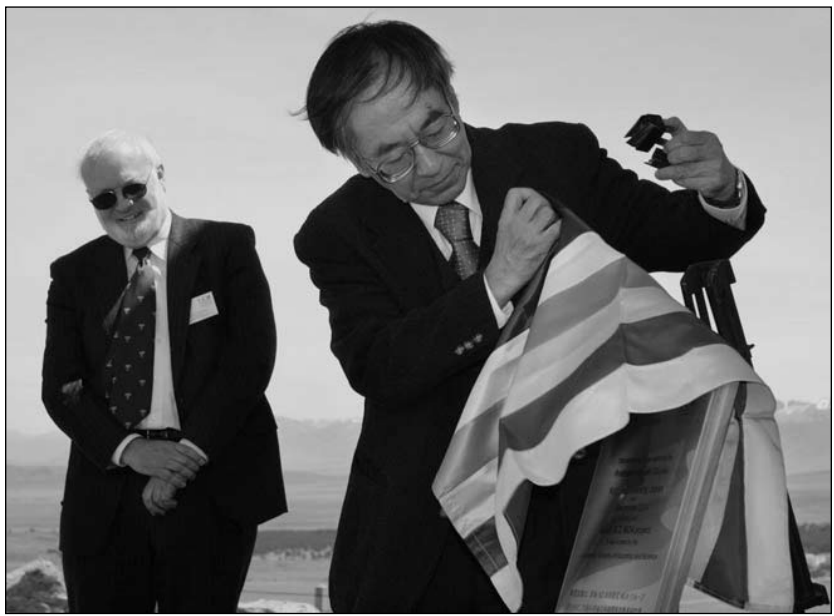

*Figure 11. Yasushi Muraki unveiling a plaque in his honour at the opening of the 1.8 m MOA telescope on 1 December 2004 with Vice-Chancellor Professor Roy Sharp of the University of Canterbury looking on approvingly.*